\documentclass[conference]{IEEEtran}
\usepackage{graphicx}
\usepackage{algorithm}
\usepackage{algorithmic}

\newcommand{\dset}[1]{\textsf{#1}}

\begin{document}
\title{End-Host Distribution in Application-Layer Multicast: Main Issues and Solutions}

\author{\IEEEauthorblockN{Genge B\'{e}la and Haller Piroska}
\IEEEauthorblockA{Department of Electrical Engineering\\
``Petru Maior'' University of T\^{a}rgu Mure\c{s}\\
T\^{a}rgu Mure\c{s}, Mure\c{s}, Romania, 540088\\
\{bgenge,phaller\}@engineering.upm.ro}}

\maketitle

\begin{abstract}

Application-layer multicast implements the multicast functionality at the application layer. The main goal of application-layer multicast is to construct and maintain efficient distribution structures between end-hosts. In this paper we focus on the implementation of an application-layer multicast distribution algorithm. We observe that the total time required to measure network latency over TCP is influenced dramatically by the TCP connection time. We argue that end-host distribution is not only influenced by the quality of network links but also by the time required to make connections between nodes. We provide several solutions to decrease the total end-host distribution time.\newline

\end{abstract}

\IEEEpeerreviewmaketitle

\section{Introduction}
For several years now group communications have been receiving significant attention from both the industry and scientific communities \cite{Scalability04, Threat08}. The main goal of group communication is to enable the exchange of information between group members that can be located across the entire globe.

One of the main application of group communications is in the field of \emph{multicast}. Historically speaking, the first multicast applications were implemented over the IP layer, also known as \emph{IP multicast} \cite{Multicast90}. However, after nearly a decade of research in the field of IP multicast, it was never fully adopted because of several technical and administrative issues \cite{Deployment00}.

Later, there have been several proposals for other multicast implementations that would be easier to deploy over the already existing and well-established Internet protocols and would require little or no modifications in existing routers. Such a survey of existing solutions was provided by El-Sayed et al \cite{Survey03}.

One of the directions that has been clearly adopted over the last few years is \emph{application-layer multicast}, which implements the multicast functionality at the application layer. The main goal of application-layer multicast is to construct and maintain efficient distribution structures between \emph{end-hosts}. These structures are constructed using an \emph{overlay} network providing the necessary infrastructure for data transfer between end-hosts.

Today's research focuses on the many aspects of application-layer multicast, including construction of overlay networks \cite{HostBased01,SelfOrganized09}, optimization issues \cite{Polynomial05} or security \cite{Security08}. In our previous work \cite{OptimalSv05} we have addressed the problem of optimally distributing end-hosts (i.e. EH) to overlay network hosts (i.e. OH) in order to minimize network latency and to distribute the load of OH. Based on a heuristic algorithm we proved that the algorithm ensures a local optimal distribution of EH in real time and thus can be used to provide a feasible solution to the distribution problem.

In this paper we focus on the actual deployment of the algorithm proposed in our previous work in a real and globally-scaled distributed system: \emph{PlanetLab} \cite{PlanetLab04}. \emph{PlanetLab} is a ``geographically distributed overlay network designed to support the deployment and evaluation of planetary-scale network services'' \cite{PlanetLab04}. Using PlanetLab, researchers can test their algorithms and systems in a real environment where nodes can become unreachable, network bandwidth can fluctuate and node processing capabilities can drop dramatically.

In order to test the real applicability of our previously proposed algorithm we have developed an overlay network in PlanetLab where nodes are connected in a complete graph model. There are several advantages for using such a graph model. First, there is no need for implementing complex routing algorithms \cite{DistrMulticast07}, which greatly simplifies the implementation and functionality of the overlay. Second, maintaining routing tables is not more complex than maintaining connections with all the other nodes. As a downside of this topology, there is a large number of connections that must be maintained, which grows exponentially with the number of OH. However, the simplicity of the routing algorithms between OH makes this topology a great candidate for using it as a leaf component in hierarchical topologies \cite{Hierarchical05, Hierarchical09}.

Existing research \cite{HostBased01, SelfOrganized09, AppLevel01} focuses on measuring the delay between nodes after the overlay has been constructed or measuring the overlay construction time after TCP connections are done. In deploying our algorithm we have observed that the total time required to measure network latency over TCP is influenced dramatically by the TCP connection time. In this paper we also argue that end-host distribution is not only influenced by the quality of network links but also by the time required to make connections between nodes.

The paper is structured as follows. In Section \ref{SEC:problem} we provide an overall presentation of the overlay network, we discuss our previous work and we identify the main problems for deploying the previously proposed algorithm. In Section \ref{SEC:measurements} we present the measurement results that were done with nodes spread across 23 countries and we provide 3 solutions for improving the performance of the measurements. Finally, we conclude with an overview of the proposed solutions and we mention some future solutions that could also be implemented.

\section{Problem Statement}
\label{SEC:problem}

\begin{figure}[htb]
\begin{center}
\scalebox{0.45}{\includegraphics{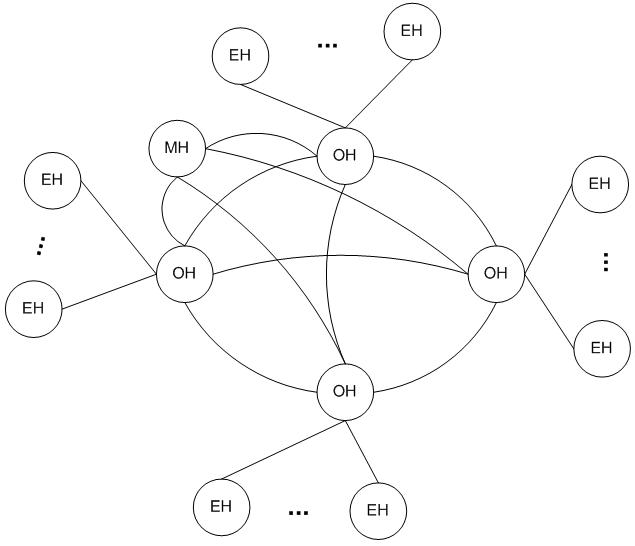}}
\caption{Multicast topology}\label{FIG:topology}
\end{center}
\end{figure}

The measurements that follow in the next sections are based on a complete graph overlay topology where EH are distributed using an heuristic algorithm. An example of such a topology is given in figure Fig. \ref{FIG:topology}, where we have illustrated the presence of 3 host types:

\begin{itemize}
  \item End-hosts (i.e. EH);
  \item Overlay-hosts (i.e. OH);
  \item Monitor-hosts (i.e. MH ).
\end{itemize}

EH are the producers and consumers of data transferred by the overlay containing the OH. MH are used to monitor the load of each OH and to distribute the connection of EH. The heuristic algorithm we proposed in our previous work is used to distribute EH to OH in order to minimize latency and to distribute the load of OH.

The distribution algorithm uses the measured latency between all OH pairs, the load of each OH and the measured latency between each EH and OH pairs. The algorithm is run by the MH each time a new EH must be connected. At this time, the EH must provide the MH its measurement results on the network latency it recorded to each OH. Based on this data and the reported load received from each OH, the MH runs the distribution algorithm.

As mentioned in our previous work, after all data is available, the algorithm executes very fast. For instance, from the simulations we run, for 100 OH the algorithm execution time for distributing a single EH is about 3.7 ms. This execution time provides a real-time applicability of the proposed algorithm.

We have chosen to deploy the proposed multicast in PlanetLab because it provides globally-available network services that can be used to run any application type that can run on a Linux OS. From the beginning of the implementation process we had to deal with several problems. First of all, network connections between PlanetLab nodes or even node CPUs can be heavily loaded, sometimes even leading to SYN\_ACK timeouts for TCP connections. Second, nodes can be rebooted at anytime by PlanetLab Central coordinators in order to ensure a software update, for software maintenance or simply because of some hardware problems. These problems must be handled by the MH in order to ensure that EH are not distributed to such nodes and that already distributed EH nodes are redistributed if necessary (i.e. on OH failure).

We also encountered several problems on the EH side. The proposed algorithm heavily relies on the measurement data provided by EH. This means that when joining the network, all EH must first measure the latency with all OH and then send this data to MH. The problem with this approach is that in some cases the response time from OH is very long, in the order of seconds as shown in the next sections. This leads to an overall distribution time in the order of seconds or even minutes, which is unacceptable.

\section{Measurement Issues and Solutions}
\label{SEC:measurements}
\subsection{Overlay Construction Time}

Although the construction of the overlay is done only once, we consider that measuring the construction time can provide useful perspective of the time required to re-construct the overlay in possible future developments. The constructing of the overlay network is not made instantly. In order to evaluate the performance and the general usability of the proposed overlay, we have measured the time needed to construct the complete graph between the overlay nodes.

\begin{table}
\centering \caption{Country and OH node count}
\label{TAB:OHNODES}       
\begin{tabular}{|l|c||l|c|}
\hline
\textbf{Country}    &  \textbf{Node count} & \textbf{Country} & \textbf{Node count} \\
\hline
\hline
Austria & 1 & Italy & 6         \\
Canada & 2 & Korea & 2          \\
France & 4 & Poland & 3         \\
Germany & 9 & Romania & 2       \\
Greece & 1 & Spain & 2          \\
Hungary & 1 & Switzerland & 1   \\
Israel & 1 & US & 5             \\
\hline
\hline
\end{tabular}
\end{table}

Deploying and starting applications on PlanetLab nodes can be done automatically using applications such as \emph{multicopy} or \emph{multiquery} that are part of the CoDeploy project \cite{Deploying04}. These allow a parallel deployment and execution of commands on a set of nodes. We have considered 5 settings with a different number of OH nodes. The OH applications were deployed on nodes from 14 countries (for the maximum number of 40 OH nodes), as shown in Table \ref{TAB:OHNODES}. After starting the OH applications, each OH connects to all other OH according to Alg. \ref{ALG:COMPLGRAPH}, where $\dset{OH}$ corresponds to the set of OH, $\dset{Cout}$ is the set of outgoing connections and $\dset{Cin}$ is the set of incoming connections.

\begin{algorithm}
\caption{Complete connections for one OH} \label{ALG:COMPLGRAPH}
\begin{algorithmic}
\STATE Let $t_1$ = @Get\_curr\_time()
\STATE Let $\dset{Cout}=\phi$

\STATE
\STATE \COMMENT{Start connection sequences}
\FORALL
{$oh\in\dset{OH}$}
\STATE $c$ = @Start\_conn\_sequence($oh$)
\STATE $\dset{Cout}=\dset{Cout}\cup \{c\}$
\ENDFOR

\STATE
\STATE \COMMENT{Wait for completion}
\STATE @Wait\_for\_completion( $\dset{Cout}$ )

\STATE
\STATE \COMMENT{Now eliminate duplicate connections}
\STATE Let $\dset{Cin}$ = @Get\_incoming\_connections()
\FORALL
{$c\in\dset{Cout}$}
    \IF {$\exists c' \in \dset{Cin}:$@Src\_address($c'$)=@Dest\_address($c$)}
    \STATE $(Meas_{out},Meas_{in})$=@Run\_measurements($c,c'$)
        \IF {$Meas_{out} < Meas_{in}$}
            \STATE @End\_connection($c$)
            \STATE $\dset{Cout}=\dset{Cout}\setminus\{c\}$
        \ENDIF
    \ENDIF
\ENDFOR

\STATE
\STATE \COMMENT{Calculate complete connection time}
\STATE Let $t_2$ = @Get\_curr\_time()
\STATE Let $G_{Time}=t_2-t_1$
\end{algorithmic}
\end{algorithm}
\begin{figure}[htb]
\begin{center}
\scalebox{0.42}{\includegraphics{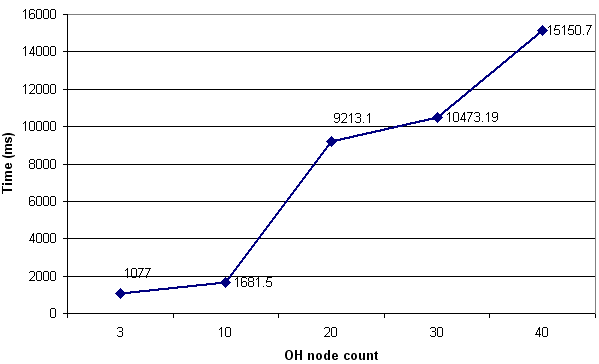}}
\caption{Complete graph construction time}\label{FIG:compgraph}
\end{center}
\end{figure}

At first, each OH starts the connection process to other OH nodes. Then, it waits for the connection process to complete. This process leads to duplicate connections between each OH node pair. In order to eliminate duplicate connections we measure the connection latency in each direction by sending a single package of 1500Bytes and we eliminate the connection with the maximum latency.

According to Alg. \ref{ALG:COMPLGRAPH}, each OH calculates a complete connection time $G_{Time}$. The complete graph construction time is the maximum of these values, as shown in Fig. \ref{FIG:compgraph}. As we can see in Fig. \ref{FIG:compgraph} the construction of the overlay is greatly influenced by the number of nodes. However, the variation is not linear because the overlay also depends on other factors such as the quality of network connections and the load of nodes. The result shown in Fig. \ref{FIG:compgraph} has the following explanation. In the first OH set (i.e. 3 nodes), all 3 nodes are located in European countries, with a minimum load. In the next OH set (i.e. 10 nodes) we have added additional nodes from Europe, one node from the US and one node from Asia. This almost doubled the graph construction time because the node from Asia was heavily loaded, with the CPU running at over 80\% almost all the time. In the next set (i.e. 20 nodes) we have added additional nodes from Asia, Canada and Europe which, because of network connection latencies and heavily loaded nodes (i.e. from Israel and Germany) has led to a quadruple time. In the next two sets (i.e. 30 and 40 nodes) we have added additional nodes from Europe and US, leading to the results shown in Fig. \ref{FIG:compgraph}.

\subsection{EH Connection Measurement Issues}
When EH nodes are started, each node first connects to all OH nodes in order to measure the network latency. The measured values are then sent to the MH that applies the heuristic algorithm developed in our previous work \cite{OptimalSv05} to determine the OH node where each EH must connect. We have identified two components that significantly influence the measured values: connection time and network latency. Let $\dset{EH}$ be the set of EH. Then, the total measurement time $M_i$ needed to be executed by an EH is:

\begin{equation}
    M_i=\max_{oh_j}\{Conn(eh_i,oh_j)+
                CummLat(eh_i,oh_j)\}
\end{equation}
where $eh_i\in\dset{EH}$, $i=\overline{1,|\dset{EH}|}$ and $oh_j\in\dset{OH}$, $j=\overline{1,|\dset{OH}|}$. $Conn$ denotes the time needed to establish a connection between $eh_i$ and $oh_j$. $CummLat$ denotes the cumulated round-trip latency calculated by measuring the time difference between sent and received packages:

\begin{eqnarray}
    CummLat(eh_i,oh_j) & = & Lat_1(eh_i,oh_j) + \nonumber \\
        && Lat_2(eh_i,oh_j) + \nonumber \\
        && Lat_3(eh_i,oh_j)
\end{eqnarray}
where $Lat_1$, $Lat_2$ and $Lat_3$ denote the round-trip latency of 3 packages.

We have considered several scenarios, with EH count ranging from 10 to 1000. EH nodes were deployed on nodes from 23 countries (for the maximum number of 1000 EH nodes), as shown in Table \ref{TAB:EHNODES}.

\begin{table}
\centering \caption{Country and EH node count}
\label{TAB:EHNODES}       
\begin{tabular}{|l|c||l|c|}
\hline
\textbf{Country}    &  \textbf{Node count} & \textbf{Country} & \textbf{Node count} \\
\hline
\hline
Argentina & 10 & Japan & 10         \\
Australia & 10 & Korea & 20           \\
Austria & 40 & Netherlands & 20         \\
Belgium & 20 & Poland & 40       \\
Canada & 100 & Portugal & 10          \\
China & 20 & Romania & 20   \\
Finland & 10 & Russia & 20             \\
France & 110 & Spain & 40             \\
Germany & 160 & Switzerland & 10             \\
Greece & 10 & Taiwan & 10             \\
Hungary & 20 & US & 240             \\
Italy & 60 & &             \\
\hline
\hline
\end{tabular}
\end{table}
\begin{figure}[htb]
\begin{center}
\scalebox{0.42}{\includegraphics{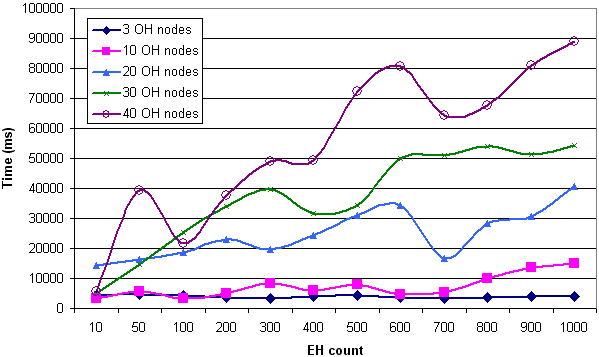}}
\caption{Average EH measurement time}\label{FIG:ehmeas}
\end{center}
\end{figure}

Each EH calculates its own $M_i$ value that is sent to the MH that calculates an average measurement time, illustrated in Fig. \ref{FIG:ehmeas}. We can see that the number of OH nodes clearly influences the overall measurement time. There are several values that break the linear trajectory. For instance, in the case of 40 OH nodes, when running 50 EH nodes the average time is 39382ms and when running 100 EH nodes the average time is reduced to 21571ms. The explanation for this behavior lies in the way that the measurements were done. Because PlanetLab offers a set of resources over the Internet that are shared among researchers, time measurements can change dramatically from one execution to another. Moreover, the measurements we made span across 10 days. We have actually seen that in one day a given node can be extremely loaded because other researchers may also be running experiments, and the next day the node can show a minimum load. This is in fact the expected behavior of nodes running in a real networking environment that greatly differs from the controlled laboratory environments.

\begin{figure}[htb]
\begin{center}
\scalebox{0.42}{\includegraphics{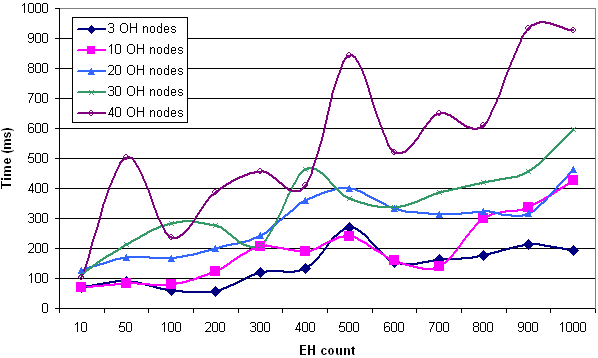}}
\caption{Average EH-OH measured latency}\label{FIG:latency}
\end{center}
\end{figure}

The values shown in Fig. \ref{FIG:ehmeas} include both the connection time and the network latency. However, as we can see from Fig. \ref{FIG:latency}, the latency is only a small part of the measurement time, with average values ranging from 68.59ms to 925.86ms.

\begin{figure}[htb]
\begin{center}
\scalebox{0.42}{\includegraphics{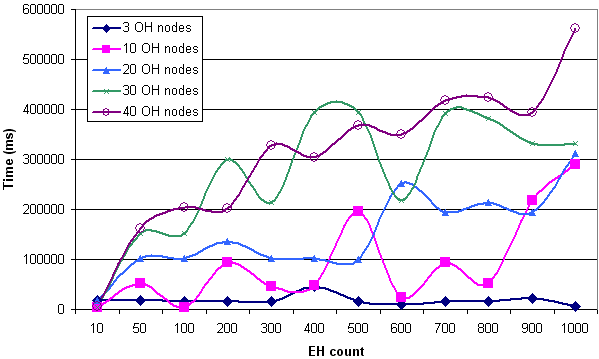}}
\caption{Maximum EH measurement time}\label{FIG:ehmeas_max}
\end{center}
\end{figure}

The values shown in Fig. \ref{FIG:ehmeas} clearly show that we should improve the performance of the measuring algorithm. At this stage, the average time needed to measure the network latency for 1000 EH nodes in the 40 OH node setting is 89000ms, which corresponds to almost 1.5 minutes. However, this is the average time, which is much smaller than the maximum time needed for an EH to make the measurements. The maximum measurement time is shown in Fig. \ref{FIG:ehmeas_max}, where we can see that the maximum time needed to make the measurements is in fact 561192ms, which is almost 9.5 minutes. The values from Fig. \ref{FIG:ehmeas_max} show that the time needed for all nodes to make the measurements are influenced by the number of OH and by the number of EH, leading to the value of 9.5 minutes, which is unacceptable.

\begin{figure}[htb]
\begin{center}
\scalebox{0.42}{\includegraphics{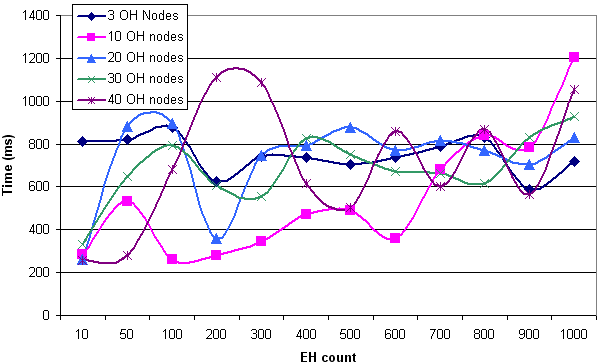}}
\caption{Average MH response time}\label{FIG:mhresp}
\end{center}
\end{figure}

The total accessing distribution time of EH is also influenced by the response time from the MH. In all our measurements the MH resides on a single node from Romania. In Fig. \ref{FIG:mhresp} we can see the average response time from the MH. Interestingly, the response time is not influenced by the number of OH or by the number of EH, but by the number of simultaneous requests that are received. EH nodes connect to MH only after completing the measurements, this is why when a large number of EH connect simultaneously to the MH we get the peaks from the figure. From the measurements we have also seen that after receiving the measurement data the distribution algorithm is running under 1ms for each request, thus the values shown in Fig. \ref{FIG:mhresp} are given by message processing and network delay.

After an EH successfully connects to the OH, it can stay connected for an unlimited time. However, if the connection is interrupted, it will reconnect to the designated OH. If the designated OH is no longer available, it must execute the measurement and distribution all over again. In case of new EH nodes, these are distributed by the MH without redistributing the already connected EH nodes.

As mentioned earlier, in case of OH failure, disconnected EH nodes initiate a new measurement and distribution process. However, in case of network failures between OH nodes, a reconnect mechanism is activated for each OH node that tries to re-establish connection with all other OH nodes, effectively trying to reconstruct the overlay.

\subsection{EH Connection Measurement Solutions}
As illustrated in the previous section, making network measurements at the application layer is mainly influenced by the connection time between nodes. The network latency factor, as opposed to the connection time, has a minimum impact on the  total time.

When EH use the proposed overlay, their main goal is not to make measurements but to actually use it to effectively distribute data. The time needed to make the measurements should thus be reduced to a minimum possible.

In this section we propose 3 solutions to the measurement problem. After implementing them, we have repeated the measurements for the 1000 EH setup, where the modifications would have a greater impact.

\begin{figure}[htb]
\begin{center}
\scalebox{0.42}{\includegraphics{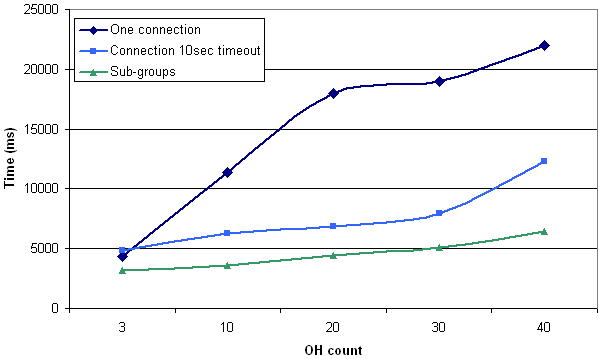}}
\caption{Average improved EH measurement time for 1000 EH}\label{FIG:solution}
\end{center}
\end{figure}

The first solution involves reducing the reconnect process count to 0, meaning that if a connect attempt fails, the EH removes the OH from its list. EH nodes usually try to connect over and over again to OH nodes until successful. This process dramatically increases the overall measurement time, as shown in the previous section. By eliminating the reconnections, we are in fact eliminating OH that are overloaded or to which we have a poor connection. The improvements can be immediately seen, as shown in Fig. \ref{FIG:solution}. In this case, for the maximum setting, with 40 OH nodes, the average measurement time drops from 89000ms to 22027ms, improving the overall measurement 4 times.

The problem with the first solution is that a connection must be timed out by the OS to eliminate the OH from the solution. As a second solution we propose an application-controlled connection timeout, opposed to network OS timeout. In this case we timed out connections that exceeded 10 seconds, decreasing the average measurement time from 89000ms to 12284ms and improving the overall measurement 7 times, as shown in Fig. \ref{FIG:solution}. The 10 seconds were chosen based on the observation that a lower timeout leads to an increased number of OH nodes eliminated from the solution. This problem is discussed in detail later in this section.

The third solution involves partitioning the OH and EH nodes into sub-groups, thus reducing the total number of OH/EH and the total number of EH/OH. The partitioning can be seen in Table \ref{TAB:PART}. As we can see from Fig. \ref{FIG:solution}, the average time required for measurements is reduced to 6459ms for 40 OH nodes, improving the overall measurement time over 13 times.

\begin{table}
\centering \caption{Sub-group partitioning}
\label{TAB:PART}       
\begin{tabular}{|l|c||l|c|c|c|}
\hline
\textbf{Sub-Group}    &  \textbf{3 OH} & \textbf{10 OH} & \textbf{20 OH} & \textbf{30 OH} & \textbf{40 OH} \\
               &  1OH/EH & 2OH/EH & 4OH/EH & 6OH/EH & 8OH/EH\\
\hline
\hline
Grp1 & 333 EH & 200 EH & 200 EH & 200 EH & 200 EH  \\
Grp2 & 333 EH & 200 EH & 200 EH & 200 EH & 200 EH  \\
Grp3 & 333 EH & 200 EH & 200 EH & 200 EH & 200 EH  \\
Grp4 &  -     & 200 EH & 200 EH & 200 EH & 200 EH  \\
Grp5 &  -     & 200 EH & 200 EH & 200 EH & 200 EH  \\
\hline
\hline
\end{tabular}
\end{table}

\begin{figure}[htb]
\begin{center}
\scalebox{0.42}{\includegraphics{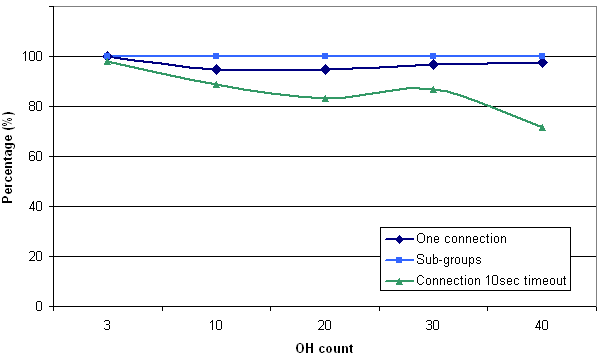}}
\caption{Average percentage of measured connections} \label{FIG:percentage}
\end{center}
\end{figure}

The direct effect of the first two solutions is that the number of OH nodes for which EH nodes test the connection reduces significantly with the reduction of the timeout. For instance, by using the OS timeout, which can range from a few seconds to a few minutes we have less eliminated OH nodes than using a fixed timeout of 10 seconds, as shown in Fig. \ref{FIG:percentage}. In case of only one connection (i.e. OS timeout) the tested percentage is 100\% for 3 OH nodes, however, this drops to 95\% for 10 and 20 nodes and then rises to 96.66\% for 30 nodes and to 97.43\% for 40 nodes. In case of application-layer timeout we have a 98.1\% for 3 OH nodes which drops to 71.79\% for 40 OH nodes.

Although the partitioning-based solution provides the best timings, it can limit sub-groups to a set of OH nodes that may not provide the optimal solution for the entire group. While the application-layer timeout mechanism seems to be the next best approach, care must be taken in choosing the timeout value because a larger connection-time does not necessarily mean that the specific node is heavily loaded, but several other factors can also influence this value, such as a momentarily busy OS, or a momentarily busy application.

Other solutions could also be applied, such as using UDP for determining the network latency between EH and OH. Such a solution would eliminate the overhead given by TCP connection. However, because the overlay uses TCP for forwarding data, making measurements by connecting to OH nodes via TCP provides a more precise view on the future behavior of OH nodes.

\section{Conclusions and Future Work}
\label{SEC:conclusions}

We presented several issues and solutions for deploying application-layer overlay networks. Based on our measurements conducted over PlanetLab, a real network testing platform, we have concluded that distributing EH nodes can not be based only on the measured network latency, but must also include other elements such as connection time or EH geographical location to reduce the time required to make the actual latency measurements.

The identified problems have several solutions. In this paper we have proposed 3 such solutions: a first one that eliminates reconnections, a second one that uses application-layer timeouts and a third one that constructs sub-groups for reducing the number of OH/EH and EH/OH. By using these solutions we have shown that the measurement time can be reduced up to 13 times for 1000 EH and 40 OH.

As future work, we intend to use UDP for the initial measurements. However, special care must be taken because a lower timing for UDP packages does not necessarily imply lower timings for TCP packages. A study must be made to determine the correspondence between UDP and TCP timings and how could UDP-based measurements be used to forecast the overhead introduced by TCP connections. This study must also take into consideration UDP packet losses that may also influence the total measurement time.


\end{document}